\documentclass[a4paper,pdftex,preprint,superscriptaddress,prb,showkeys]{revtex4} 
\pdfoutput=1

\usepackage{graphicx} 
\usepackage{threeparttable}
\usepackage{amssymb,amsmath}
\usepackage{multirow}
\usepackage{rotating}
\usepackage{float}
\usepackage{units}
\usepackage{textcomp}
\usepackage{color}

\usepackage[normalem]{ulem}
\usepackage{dcolumn}
\usepackage{bm}

\def\be{\begin{eqnarray}}
\def\ee{\end{eqnarray}}
\def\r{{\bf r}}

\def\E{{\bf E}}
\def\H{{\bf H}}

\def\F{{\bf F}}
\def\p{{\bf p}}
\def\m{{\bf m}}

\def\G{{\tensor{\bf G}}}

\def\alfa{\tilde{\alpha}}

\begin{document}

\title{Controlling dispersion forces between 
small particles with \\ artificially created random light fields}

\author{Georges Br\"{u}gger}
\author{Luis S. Froufe-P\'erez}
\author{Frank Scheffold}\thanks{Equal senior authorship. \\ Correspondence and requests for materials should be addressed to F.S. (email: frank.scheffold@unifr.ch) or J.J.S. (email: juanjo.saenz@uam.es).}
\affiliation{Department of Physics, University of Fribourg, Chemin du Mus\'{e}e 3, CH-1700, Fribourg, Switzerland}

\author{Juan Jos\'e  S\'{a}enz}\thanks{Equal senior authorship. \\ Correspondence and requests for materials should be addressed to F.S. (email: frank.scheffold@unifr.ch) or J.J.S. (email: juanjo.saenz@uam.es).}
\affiliation{Depto. de F\'{i}sica de la Materia Condensada, Instituto Nicol\'{a}s Cabrera and Condensed Matter Physics Center (IFIMAC), Universidad Aut\'{o}noma de Madrid, 28049 Madrid, Spain}
\affiliation{Donostia
International Physics Center (DIPC), Paseo Manuel Lardizabal 4, 20018
Donostia-San Sebastian, Spain}

\date{\today}

\begin{abstract}
{\bf Appropriate combinations of laser beams can be used to trap and manipulate small particles with ``optical tweezers'' 
as well as to  induce  significant ``optical binding'' forces between particles. 
These interaction forces are usually strongly anisotropic  depending on the  interference landscape of the external fields. 
This is in contrast with the familiar isotropic, translationally invariant, van der Waals and, in general, Casimir-Lifshitz interactions between neutral bodies arising from random electromagnetic waves generated by equilibrium quantum and thermal fluctuations. 
Here we show, both theoretically and experimentally, that dispersion forces between 
small colloidal particles can also be induced and controlled using artificially created fluctuating light fields. Using optical tweezers as gauge, we present experimental evidence for the predicted isotropic attractive interactions between dielectric microspheres induced by laser-generated, random light fields. These light induced interactions open a path towards the control of translationally invariant interactions with tuneable strength and range in colloidal systems. } \end{abstract}

\keywords{optical forces, waves in random media, field-field correlations, Casimir effect, colloidal interactions}
\maketitle
\tableofcontents

\section{Introduction}
 The familiar isotropic dispersion forces between neutral objects arise from random electromagnetic waves generated by equilibrium quantum and thermal fluctuations \cite{Israelachvili1991,Parsegian2006,Milton2011}. Depending on the context, these forces are known as non-retarded van der Waals-London, Casimir-Lifhsitz  and, more generally, Casimir forces \cite{Israelachvili1991,Parsegian2006,Milton2011,Rodriguez2011}.  
%
 The interplay between  Casimir forces and electrical double layer forces, which forms the basis of the famous DLVO theory \cite{Israelachvili1991} describing the forces between charged surfaces in a liquid medium,  plays a key role in the colloidal behavior observed in biological fluids (e.g. proteins, biopolymers, blood cells), foodstuffs (e.g. dairy, thickeners, emulsions and creams) or suspensions (e.g. pharmaceuticals, slurries, paints, inks) \cite{liang2007interaction}.
Colloids have also been shown  to be extremely well suited for the study of phenomena such as crystallisation, the glass transition, fractal aggregation and solid-liquid coexistence \cite{frenkel2002playing,poon2002physics,pham2002multiple}. 
 External  control of  isotropic interactions  in colloidal systems is therefore of key importance. Temperature sensitive swelling of 'smart' microgel particles offers control over soft repulsive forces but the process is slow, shows hysteresis \cite{pelton2000temperature} and 
 the properties of the colloids are altered while swelling. In some cases magnetic and dielectric dipolar forces can be induced by external fields but these interactions are strongly anisotropic leading to the formation of chains or anisotropic domains \cite{yethiraj2003colloidal}.  Other ways to control colloidal interactions usually involve the change of composition: Adding and removing electrolytes the range of electrostatic repulsions can be tuned and, by dissolving macromolecules of appropriate size, attractive depletion forces can be induced \cite{poon2002physics,pham2002multiple}. However, despite their widespread and successful use, these strategies are still tedious and slow and do not provide the level of control over interaction forces that, as discussed here,  could be achieved by using external laser fields. 
 
Intense light fields can be used to trap and manipulate small particles \cite{Ashkin1986,Burns1990,mikhael2008archimedean} as well as to  induce  significant optical binding forces \cite{Thirunamachandran1980,Burns1990,Dholakia2010} which, in general, are not translationally invariant, showing  a strong anisotropy that depends on the  interference landscape of the external fields \cite{Dholakia2010}.   Here we  show that
 artificially generated random fields with appropriate spectral distribution can provide control over attractive and repulsive isotropic (and translationally invariant)  interactions with tuneable strength and range.   In contrast with Casimir interactions, where the forces are dominated by the material's response at low frequencies,  our results open a new way to explore the peculiar optical dispersion of small particles and artificial metamaterials by selecting the spectral range of the random field.  As an example, we predict that
the interactions between semiconductor particles  with relatively  high refractive index, can be 
tuned from attractive to  strongly repulsive when the external frequency is tuned near the first magnetic Mie-resonance \cite{Garcia_Etxarri2011}. 

Using optical tweezers as gauge, we present experimental evidence for the predicted isotropic attractive interactions between dielectric microspheres induced by laser-generated, quasi-monochromatic random light fields. We note that isotropic optical forces between particles act instantly and can therefore also be applied dynamically. This can potentially be useful to anneal defects in periodic structures such as photonic crystals, to increase the effective temperature by optically 'shaking' particles  or to stabilize non-equillibrium phases such as supercooled liquids and, in general, to control the self-assembly and phase behaviour
of colloidal particle assemblies on nano- and mesoscopic length scales \cite{liang2007interaction,frenkel2002playing}. 


\section{Random light induced  interaction forces  between two arbitrary objects.  
}

Early work by Boyer \cite{Boyer1973} derived Casimir interactions  between small polarizable particles from classical electrodynamics  with a  homogeneous and isotropic classical random electromagnetic field  having the spectral density of quantum blackbody radiation including the zero-point radiation field. 
 Here we extend these ideas to  ``external'' artificial random fields with arbitrary spectral density, obtaining an explicit expression for the interactions between two arbitrary dielectric objects, which allows a compact description of, random light field induced, interaction forces from  dipolar (atomic or nanometer-scale) to macroscopic objects. As a limiting case, when the spectral density of the random field corresponds to that of quantum blackbody radiation, we recover the exact trace formulae for Casimir interactions between arbitrary compact objects \cite{kenneth2006,Emig2007}.  
 Related trace expressions have also been obtained to describe non-equilibrium Casimir interactions between objects held at different temperatures \cite{Messina2011,Kruger2012}.


We first analyze the connection between random light induced  interaction forces and Casimir interactions between two arbitrary objects.  
We assume that object/particle A is at the origin of coordinates and particle B is displaced a distance r along the positive $z$-axis in an otherwise transparent and non-dispersive homogeneous medium with real refractive index $n_h=\sqrt{\epsilon_h}$.  
We consider that the particles are illuminated by a quasi-monochromatic random field of frequency $\omega$ which
  can be described as a superposition of plane waves  with random phases and polarizations, propagating in all directions.
Each particle can be seen as made of discretized, $N_{A}$ and  $N_{B}$, identical cubic elements   of volume $v$ and relative permittivity $\epsilon(\omega)$, which
act as  small polarizable units  with an induced dipole proportional to the polarizing field, i.e. $
 \p(\r_n, \omega) = \epsilon_0 \epsilon_h  \ \alpha(\omega) \E_{\text{inc}}(\r_n, \omega)  $,
 where $\alpha(\omega)$ is the polarizability   given by
$
\alpha(\omega) \equiv  v \alfa_0/[1-i v k^3\alfa_0/(6\pi)]$ with $\alfa_0(\omega) \equiv 3 (\epsilon(\omega)-\epsilon_h)/(\epsilon(\omega) + 2\epsilon_h).$ 
 In the presence of a fluctuating polarizing  field, $\E_{\text{inc}}(\r,t)$, the induced dipoles
are  fluctuating quantities and 
the time averaged force on particle B along the $z$-axis may be
written as \cite{Boyer1973,Chaumet2000}
 \be
 \langle F_z \rangle_B = \sum_n^{N_B} \biggl\langle \p(\r_n^B,t) \left. \frac{\partial}{\partial z} \E_{\text{inc}}(\r,t)\right|_{\r=\r_n^B} \biggr\rangle.
 \ee 
The total force can be seen as the sum of different contributions. Although for random illumination there is no net force on an isolated particle, the scattered field by B can be reflected back by  particle A leading to a series of multiple scattering events which give rise to a net interaction force between them.  An additional contribution arises from the correlations between the induced dipoles. At a first sight, one could think that the incoming exciting  fields on the two objects will  be completely uncorrelated. 
However, in a random, statistically stationary and homogeneous  electromagnetic field, the fields at two distant points are correlated, with a cross spectral density of the correlations  identical to that of blackbody radiation \cite{Setala2003}. 
In absence of absorption (when the relative permittivity $\epsilon(\omega)$ and $\alfa_0$  are real numbers), the sum of the two contributions lead to a conservative interaction force which can be expressed in terms of the ${\bf{T}}$-matrix \cite{mishchenko2002scattering} of each individual object  [see Supplementary Material \ref{TheorSuppl}] 
 \begin{align}
 \F &= - \bm{\nabla} U(r) \\
 U(r) &=  \int_0^\infty d \omega \ 
\frac{2 \pi}{k^3} u_E(\omega) \ \text{Im}  \text{Tr} \left[\ln\left(  {\bf{I}} - \G_{B,A}{\bf{T}}_A\G_{A,B}{\bf{T}}_B \right) \right] \label{Uexact}
\end{align}
 where $ [u_E(\omega) d\omega] = U_E(\omega) $ is the energy of the fluctuating electric field per unit of volume and $k=n_h\omega/c$ is the wave number ($c$ is the speed of light in vacuum).  $\G_{B,A}$ is the Green tensor connecting the two objects and ``Tr'' stands for the trace of the ${\bf{T}}$-matrix \cite{mishchenko2002scattering}.
 The dependence of the interaction on distance is completely contained in $\G_{B,A}$ whereas all the shape and material dependence is contained in the ${\bf{T}}$ matrices.  
 The connection with Casimir interactions can be made through Boyer's approach:
When the objects are in equilibrium with a quantum blackbody radiation, the energy density $U_E(\omega) $ corresponds to the electric quantum  zero-point  fluctuations (at zero temperature and positive $\omega$) given by \cite{Landau_Statistical,Joulain2005}
 $U_E(\omega) = u_E(\omega)d\omega =  \hbar k^3/(4\pi^2) d \omega$. For absorbing (emitting) particles,
 we must include an additional contribution to the total force coming from the fluctuating dipoles  and the corresponding radiated fields \cite{henkel2002} (linked through the fluctuation-dissipation theorem). Interestingly, in equilibrium, this additional contribution conspires with the force due to the field fluctuations  to give a total interaction potential which is exactly given by  Eq. \eqref{Uexact}, now including light absorption and emission (i.e. [$\text{Im}\{\epsilon(\omega)\} \ge 0$) and  we recover the exact Casimir interaction between arbitrary compact objects \cite{kenneth2006,Emig2007}. 
 In contrast with the traditional Casimir forces, equation \eqref{Uexact} 
 opens the path towards complete control and tunability of isotropic dispersion forces between compact bodies by  tailoring the spectral density of artificially generated random fields.
 
\subsection{Interactions between dipolar electric and magnetic particles}

 In the limit of small dipolar particles, Eq. \eqref{Uexact} leads to  Renne's  result \cite{Renne1971} obtained  from quantum-electrodynamic calculations which is identical  to  Boyer's \cite{Boyer1973} based on classical electrodynamics (the more familiar Casimir-Polder result is recovered  in the weak scattering limit). Recent theoretical works on optical binding (OB) between dipolar particles under non-coherent random illumination \cite{Rodriguez2009,Sukhov2013} and on  dipolar particles near a planar fluctuating light source \cite{Aunon2013} suggested striking similarities between dipolar optical forces in random fields and Casimir interactions.   
However,  from a practical point of view, the creation of isotropic random light fields and the direct detection of the resulting weak optical binding forces between particles  at room temperature is challenging. 

For small non-absorbing particles,  the interaction energy far from resonance is attractive but always much smaller than $k_BT$ for realistic  power densities, while, at resonance, the polarizability is purely imaginary [at resonance, $\alfa_0 \rightarrow \infty$, i.e.  $\alpha^2 <0$] which leads to an effective repulsion. The latter was shown to play a key role in understanding the collective behavior of optical trapped neutral atoms \cite{Walker1990}. Plasmonic or polaritonic nano-particles show a high real and imaginary polarizability close to a resonance but this can lead to a significant  increase of the temperature \cite{seol2006gold}.  Non-absorbing semiconductor nanoparticles,  with relatively  high refractive index, or  colloidal dielectric micron sized particles would offer an attractive laboratory to verify our predictions.  Semiconductor particles, e.g. silicon spheres with index of refraction $\sim$ 3.5 and radius $\sim$ 200nm \cite{ibisate2009silicon}, present strong electric and magnetic dipolar resonances in telecom and near-infrared frequencies, (i.e. at wavelengths Å $1.2 - 2\mu$m) without spectral overlap with quadrupolar and higher order resonances \cite{Garcia_Etxarri2011}.  Assuming that the scattering  by these Si particles can  be described by just dipolar electric and magnetic fields, it is possible to obtain an exact  closed expression for the interaction potential, Eq. \eqref{Uexact}, in terms of the electric and magnetic polarizabilities [see Supplementary Material \ref{TheorSuppl}]. Figure \ref{fig:differentpotentials} (I) illustrate how  monochromatic random illumination    induces a pair potential between two identical 230nm radius silicon nanospheres which  
can be 
tuned from attractive off-resonance ($\lambda \sim 2 \mu$m  )
 to  strongly repulsive when the external wavelength is tuned near the first  Mie's magnetic resonance ($\lambda \sim 1.6 \mu$m).
 \begin{figure}[t]
\begin{center}
\includegraphics[width=10cm]{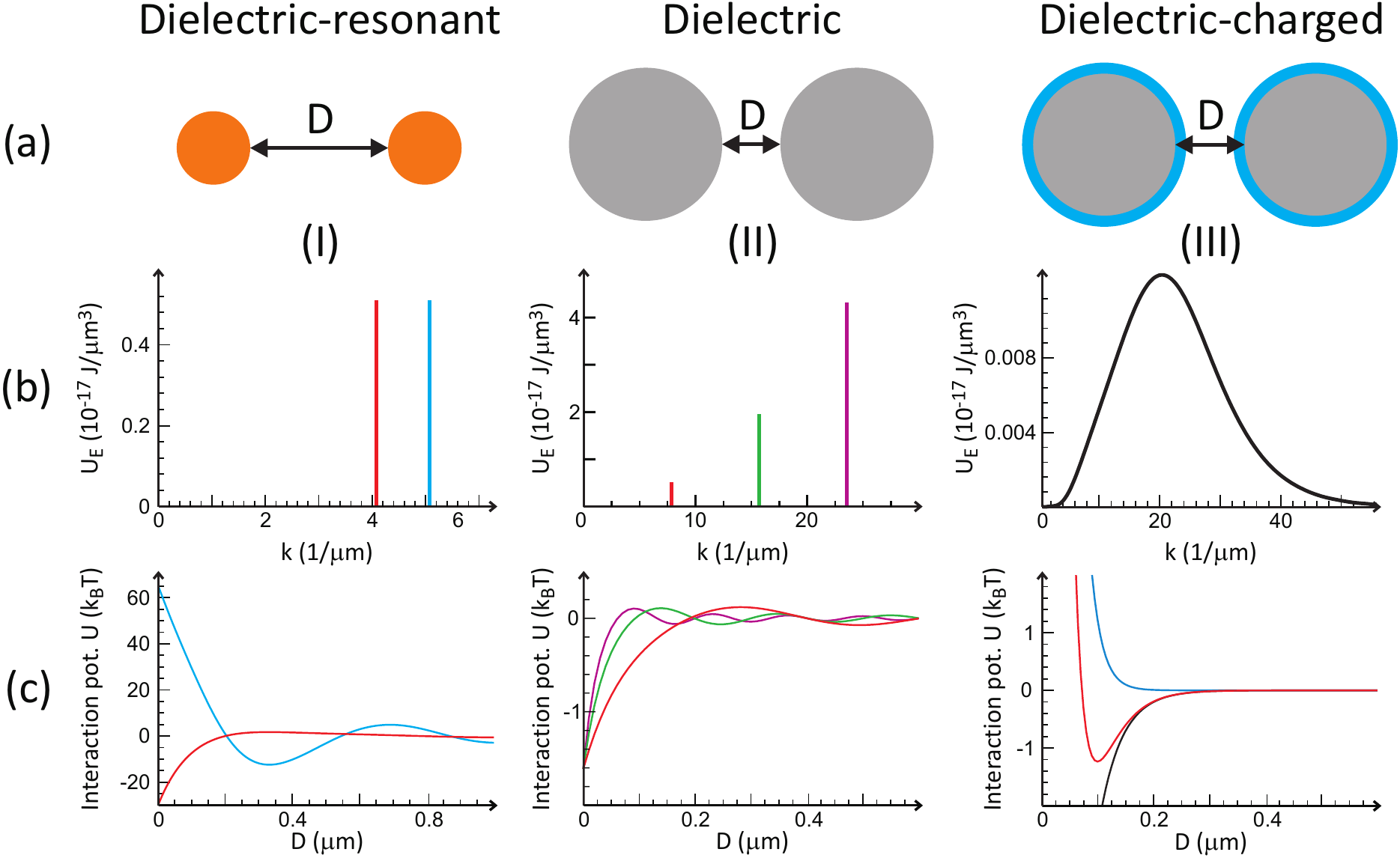}
\caption{ {\small {\bf Tunable interactions induced by random light fields}.
(a) Three representative examples of  ÕdesignerÕ random-light-induced pair potentials $U(D)$.  
(b)  Electric field energy densities $U_E(\omega)$ versus $k=n_h \omega/c$. Their corresponding interaction potentials are depicted in (c). 
(I) Exact calculation based on Eq. \eqref{Uexact} for nonabsorbing semiconductor particles where the optical response is well described by the electric and magnetic polarizabilities. Random light illumination allows for switching between strongly attractive (red) and strongly repulsive (blue) interactions. The examples show the interaction between two $R=230~$nm silicon spheres ($n=3.5$) in water ($n_h=1.33$). Based on equation~\ref{JuanjoRGDmT} we calculate in (II)  the interaction potential due to a monochromatic random light illumination between uncharged dielectric particles at different wavelengths (compare the corresponding colors). By changing the wavelength (or wave-number $k$) the range of the interaction potential can be tuned precisely. 
(III) Properly designing the energy density spectrum of random light fields allows for exponentially attractive dispersion potentials (black curve). This property can be used to produce Morse type potentials on a colloidal level using dielectric-charged particles. The exponential electrostatic double-layer repulsion calculated for a contact potential of 180~k$_B$T and a Debye length of 20~nm (blue) and the exponentially attractive dispersion interaction (black) superimpose to a colloidal Morse-potential (red). 
}}
\label{fig:differentpotentials}
\end{center}
\end{figure}
 \subsection{Interactions between micron sized dielectric particles}
 
To derive a simplified theoretical expression for the interaction potential between micron sized dielectric particles we approximate Eq.(\ref{Uexact}) at lowest order in perturbation theory  (see Supplementary Material \ref{TheorSuppl}). In this limit the interaction energy can be seen as given by  a pairwise interaction $u(|\r_B-\r_A| )$   summed over the volume of the spheres, i.e. like a Hamaker's integral \cite{Israelachvili1991}, 
 \be
  U(D)  &=&  \int_{|\r_A| \le R}  \frac{d^3\r_A }{v}   \int_{|\r_B -r\hat{\bm{e}}_z|\le R}\frac{d^3\r_B}{v} \  
 u(|\r_A-\r_B|)  
=  -\sum_{n=1}^{N}K(k_n) \, {\cal U}\left(D,R,k_n\right)  \label{JuanjoRGDmT}
\ee  with size independent coefficients $K(k_n)={2U_E(k_n)}\pi {k_n^3}\ (\alfa_0/4 \pi)^2 $ as a measure of the materials contribution to the interaction potential for a given spectral component $k_n=n_h\omega_n/c$ of the random light field. $D=r-2R$ denotes the gap distance between the surfaces and $r$ is the distance between the centers of the two spheres. $U_E\left(k_n\right)$ is the power density of the random light field for a specific wave number $k_n$. For monochromatic illumination $   U(D) =-K(k) \times {\cal U}(D,R,k)$ and in the limit of $kR \ll 1 \ll kD$ we recover the expected results for dipolar particles \cite{Thirunamachandran1980,Sukhov2013} where the interaction energy is proportional to the squared volume of the particles. Interestingly, for relatively large particle sizes (in an intermediate  regime $1 \ll kD \ll kR$) we find:
 \be
 U\left( D \right) \approx -2 U_E \alfa_0^2  \frac{\pi^3}{4 k^3} \frac{R}{D} \sin(2kD) \quad, \quad (1 \ll kD \ll kR) \label{JuanjoRGDexpand}
 \ee
 i.e. for relatively large particle sizes, our approach predicts that  the interaction energy should scale with the particle's radius.  Otherwise  Equation \eqref{JuanjoRGDmT} can be easily  computed numerically. Figure \ref{fig:differentpotentials} (II, III) illustrates the predicted theoretical sensitivity of the induced pair interactions between low-index (n=1.68) micron-sized spheres to changes in the wavelength of the illuminating random field. 
 Since forces can be strongly frequency dependent, actual dispersion forces could be manipulated by tuning the spectrum of the random light field. As an example, in Figure \ref{fig:differentpotentials} (II) we show that it is possible to tailor the range of an effective attractive dispersion interaction potential with constant depth by selecting different wavelengths and field power densities  [calculations are performed using Eq. \eqref{JuanjoRGDmT}]. Selecting an appropriate continuous spectrum of the random light fields it is also possible to design a nearly exponential attractive potential. Taking into account the well known exponential electrostatic double layer repulsion, it would theoretically be possible to induce an effective Morse-type interaction potential on a colloidal level, as shown in Fig. \ref{fig:differentpotentials} (III).

\begin{figure}[ht]
\begin{center}
\includegraphics[width=10cm]{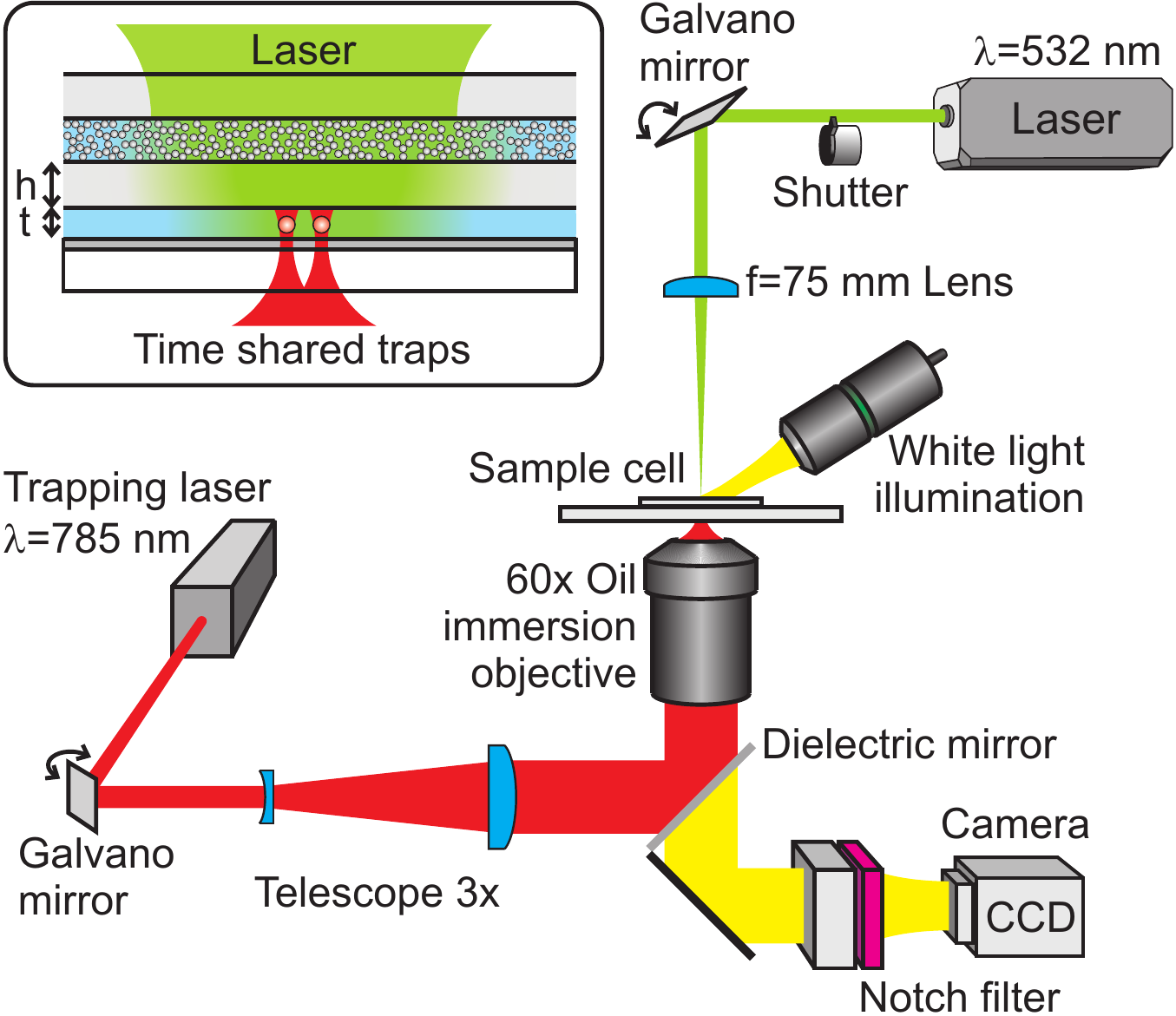}
\caption{{\small {\bf Experimental setup}. An intense green laser  ($\lambda=532$nm) is weakly focussed on one side of the sample cell using a $f=75$mm lens. A turbid scattering layer of thickness $20 \mu$m at the entrance of the cell creates a random field distribution inside a light filled cavity  of dimensions $\simeq (50\mu$m)$^3$. From the opposite side the sample cell is illuminated with a tightly focussed near infrared laser ($\lambda=785$nm) creating a set of two time shared optical traps. 
Both laser beams can be steered by galvano mirrors. A white light source is employed for the broadband illumination of the cell allowing the tracking of the particles by video microscopy. Different filters are used to spectrally isolate the different optical paths.  The interaction potential between the two trapped spheres is obtained by monitoring the thermal Brownian motion inside the optical traps with a digital camera (CCD). The inset shows an enlarged view of the sample cell: Two micron-sized colloidal particles suspended in water are trapped in the center of water filled layer of thickness $t=15 \mu$m. The water layer and the turbid layer are separated by a glass wall of thickness $h\simeq20 \mu$m. The dichroic mirror at the bottom of the clear layer reflects more than $99\%$ of the incoming light leading to multiple reflections inside the cavity.  }}
\label{fig:sample_cell}
\end{center}
\end{figure}

 \section{Experimental observation of random light induced interactions between dielectric microspheres}

Experimentally, the manipulation of relatively large, micron-sized, dielectric particles requires high intensity laser fields: In conventional optical tweezing experiments, the focussed laser intensity $I_0=P/A$ required to tightly trap a micron-sized polystyrene sphere (refractive index $n\sim1.6$), is on the order of mW/$\mu$m$^2$ and the corresponding power density is $U_E=n_h/c\cdot I_0$.  Therefore, to create random light fields with a comparable power density the surface area $A$ cannot exceed some tens of micrometers squared, for an incident laser power $P$ on the scale of Watts. To this end we have designed a miniaturised sample cell that allows for the simultaneous creation of a random light field in a small cavity by illumination with a strong green laser $\lambda=532$nm and the observation of optical binding forces between two isolated microspheres (Figure \ref{fig:sample_cell}). 
We use the technique of time shared optical tweezers\cite{Sasaki1991, Fallman1997, Mio2000} in combination with umbrella sampling\cite{Huang2009} to probe the particle pair interaction potential $U(r)$ of two melamine microspheres (n=1.68) as a function of their center-to-center separation distance $r$. The principle of this method is to trap, at the same time, two particles in two identical optical tweezers that are separated by a distance $r_0$ using a near infrared laser beam with a wavelength of 785~nm. By monitoring the relative motion of the trapped particles with a digital camera information about the pair potential $U(r)$ around $r_0$ is gained. In contrast to vacuum, the interactions between particles suspended in water commonly involve both van der Waals and screened electrostatic repulsive interactions (DLVO interactions). The latter ensures the stability against particle coagulation. Equally, in our measurements, this repulsive part dominates at very short distances and  in turn this allows us to probe the superimposed light induced attractions, without particles sticking together irreversibly.  We follow the approach of Grier and coworkers \cite{Polin2008} to obtain an autocalibrated measurement of the colloidal interaction potential by analysing the differences between the distributions of the particle positions in the presence and absence of optically induced forces (see also Supplementary Material \ref{ExpSuppl} and  \ref{ExpSuppl2}). Thus in our experiments, by turning on and off the random light field, we obtain directly the light-induced interaction potential  without an explicit measurement of the complete pair interaction potential. Because of the finite width of the laser traps the particles sample only a small range of $U$ around $r_0$. In order to determine the potential over a wider range of distances we repeat the measurements for different relative positions of the optical traps (umbrella sampling). Typically we start at an average trap separation that is near contact and increase $r_0$ in steps of $\Delta=40$~nm until 6 different trap positions are scanned. $\Delta$ is chosen to be smaller than the width of the trap potential $2\sigma\approx140$~nm to have sufficient overlap between neighbouring trap separations.

\begin{figure}[t]
\begin{center}
\includegraphics[width=8cm]{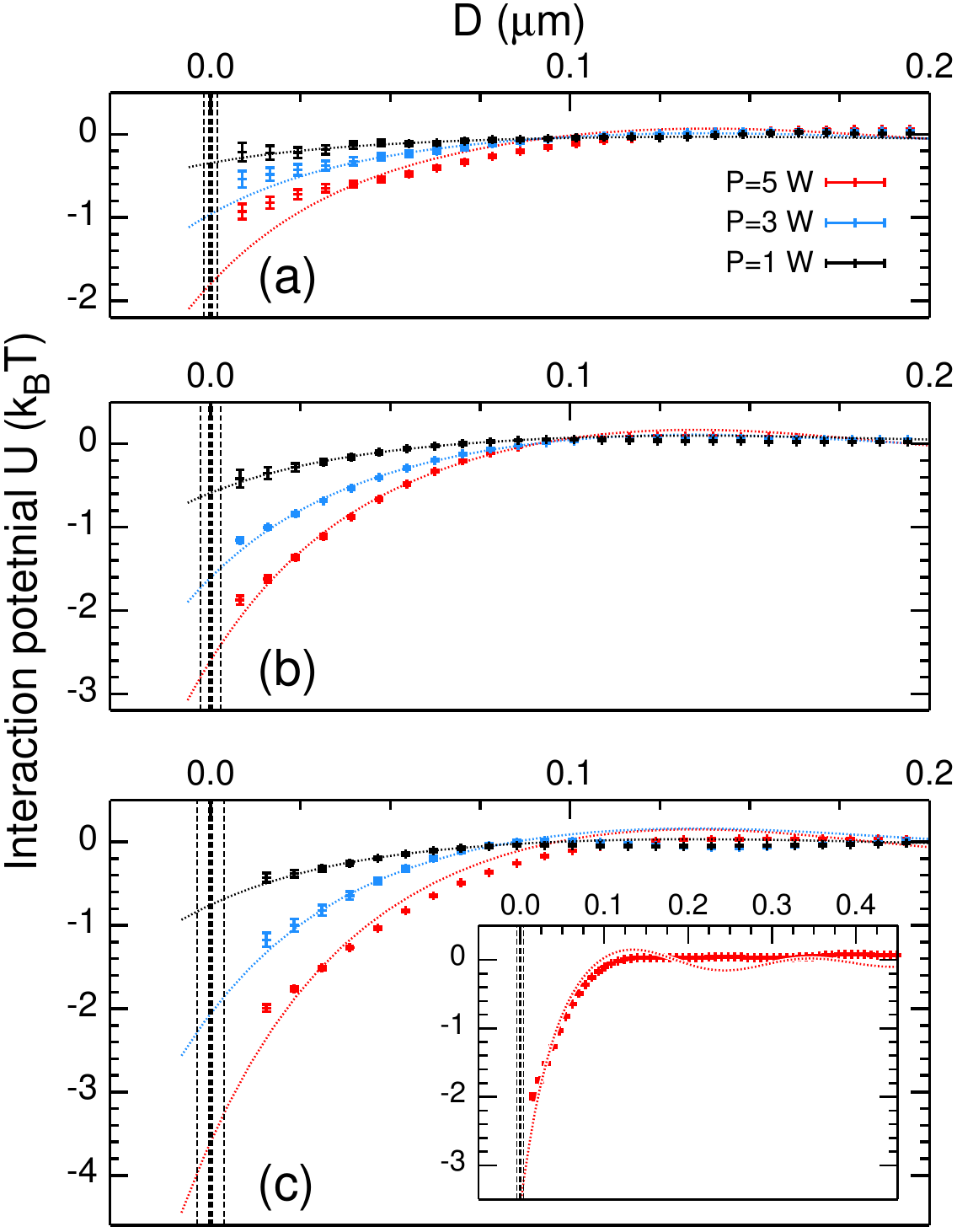}
\caption{{\small {\bf Optical binding between dielectric microspheres in random light fields}.  From the analysis of the the thermal motion in the two adjacent optical traps the particle interaction potential $U(D)$ is obtained experimentally. The melamine (n=1.68) microspheres  have sizes $2R=2~\mu$m (a), $2R=3~\mu$m (b) and $2R=4~\mu$m (c). All particles are suspended in a 2.7~mM KCl aqueous solution. In the figure the measured interaction potential is shown for three different laser power settings $P$ in each panel. The  vertical lines show the estimated contact positions and the experimental uncertainty $\Delta D\sim \pm 5$nm. By numerical evaluation of Eq.~\ref{JuanjoRGDmT} we can fit the data and then extrapolate $U(D;P)$ to obtain values for the contact potential $U_0$ as a function of particle size $2R$ and laser power $P$.  The inset shows an enlarged plot for $2R=4\mu$m and $P=5$W.}}
\label{fig:Url}
\end{center}
\end{figure}

The final results for $U(D=r-2R)$, shown in Figure \ref{fig:Url}, are obtained by averaging over 16 independent experimental runs carried out under the same conditions. We find clear evidence for optical binding in a random light field. Moreover, we are able to quantify the potential $U(D=r-2R)$  as well as its dependence on the incident light power. Varying the laser power from 5.0~W to 1.0~W the attractive potential weakens. For an estimate of the contact potential $U_0 \equiv U(D=0)$ we adjust the prefactor in Eq. \ref{JuanjoRGDmT} for a best fit to the data as shown in Figure~\ref{fig:Url}.  The overall agreement between experiments and theory is remarkable except for the predicted oscillation of the interaction potential at large inter particle distances. 
This could be partially associated to the absence of a full $4\pi$ isotropic illumination in the experiment due to the absence of incoming photons with momentum near parallel to the mirror. It could also be due to the statistical average over different experimental realizations as positions extracted from these different measurements can be slightly shifted with respect to each other, which may result in a smearing of the curves. Finally we note that the slight mismatch with theory could also be due to the approximations made when deriving Eq. \ref{JuanjoRGDmT}.

\begin{figure}[t]
\begin{center}
\includegraphics[width=10.0cm]{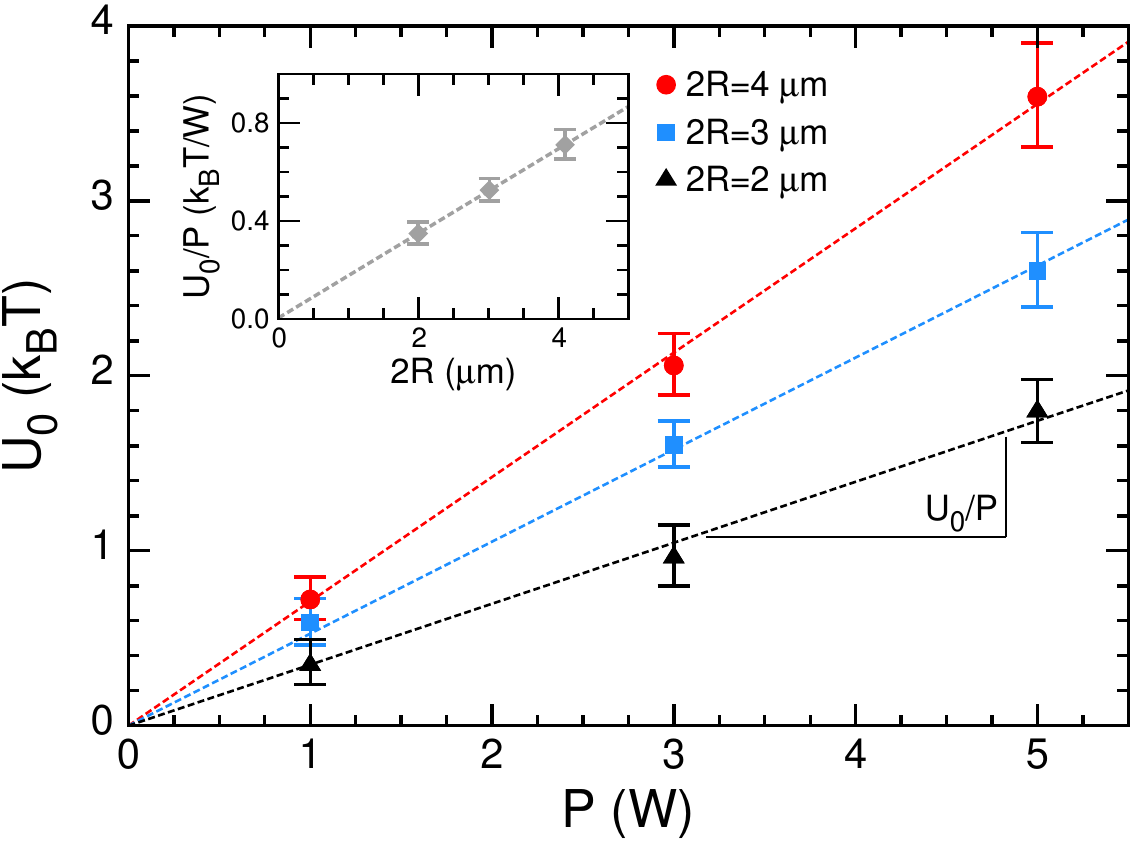}
\caption{{\small {\bf Laser power and particle size dependence of the contact potential}. Contact potential $U_0$ as a function of the laser power $P$ for particles of different size $2R=2\mu$m$,3\mu$m$, 4\mu$m.  From a linear fit (dashed lines) we estimate the normalized contact potential $U_0/P$ for the respective particle sizes. The inset shows the dependence of $U_0/P$ on the particle diameter $2R$ and the dotted line is a linear fit to the data $U_0/P \sim 0.17 \cdot k_B T/W\times (2R/\mu m)$.}}
\label{fig:master_plot}
\end{center}
\end{figure}

In Fig.~\ref{fig:master_plot} we display results for the contact potential $U_{0}$ as a function of laser power $P$ for the three different particle sizes. From the slope of the linear fits (dashed lines) we can extract the normalized contact potential $U_0/P$ which does not depend on the laser power.   $U_0/P$  increases with particle size and the data set is consistent with the linear increase as predicted by Eqns \ref{JuanjoRGDmT} and \ref{JuanjoRGDexpand}. Finally we attempt a quantitative comparison between the experimental results and the theoretical predictions for the contact potential $U_0$. We note that such a comparison is based on a number of uncertainties related to the approximations made in the theory as well as the experiment. From Eq. \ref{JuanjoRGDmT} we compute the theoretically predicted  values  for melamine microspheres with an index of refraction of $n_P=1.68$ in water ($n_h=1.33$)  as a function of particle size. For the  cross-sectional area of the light filled cavity we use $A\simeq0.004$~mm$^2$.  Moreover we take the  light power in the cavity equal to the incident laser power for a vacuum-wavelength $\lambda_0=532$~nm. Under these assumptions the theory predicts $U_0^{theory}/P \sim 0.2 \cdot k_B T/W\times (2R/\mu m)$. This results matches the experimental value $U_0^{experiment}/P \sim 0.17 \cdot k_B T/W\times (2R/\mu m)$, inset Figure \ref{fig:master_plot}. Given the approximations made such a near quantitative agreement might be somewhat fortuitous but nonetheless the overall agreement between experiment and theory supports our findings.

\section*{Acknowledgements}
This work was supported by the Swiss National Science Foundation through projects  No.132736 and No.149867 and through the National Centre of Competence in Research {\it{Bio-Inspired Materials}}. JJS acknowledges financial support by the Spanish MEC  (Grant No. FIS2012-36113), Comunidad de Madrid  (Grant No. S2009/TIC-1476 -Microseres Program-) and by an IKERBASQUE Visiting Fellowship (J.J.S.). We thank J-F. Dech\'ezelles for providing us with the PMMA particles.

 \section*{Author contribution}
  JJS and FS conceived the study. GB and FS designed the experiment. GB carried out the experiments. JJS derived the theory. GB and LF performed the numerical calculations.  GB, FS, LF and JJS  analysed and interpreted the data. JJS, FS and GB wrote the manuscript.


\clearpage

 \appendix
 
\section{Experimental Methods}
\label{ExpSuppl}
 
 A focused green laser beam (Coherent  Verdi-V5, $\lambda=532$~nm, calculated $1/e^2$ beam-waist $2w_0=25$~$\mu$m) with a laser power of up to 5 Watts is focused with a $f=75$ mm lens on the back surface of a glass capillary filled with a dense amorphous solid composed of PMMA (Polymethylmethacrylate) beads, diameter $\sim 0.4\mu$m, with a layer thickness of $\sim20~\mu$m. This first layer scatters more than $99~\%$ of the incident power and thereby creates a random light field in the sample cavity of thickness $t \simeq 15~\mu$m (inset, Figure \ref{fig:sample_cell}). The latter is filled with a very dilute suspension of melamine microspheres (Microparticles GmbH, Germany) with a diameter of $2~\mu\mathrm{m}\le  2R \le  4~\mu$m and refractive index $n=1.68$, dispersed in aqueous solution and sealed with UV-curable glue. In order to screen the electrostatic repulsions we add 2.7~mM KCl. This results in a Debye screening length of $\lambda_D\simeq6$~nm. The thickness of the layer containing the melamine microspheres is controlled by adding a small amount of $t=2R=15~\mu$m silica spheres that serve as spacers. Both layers are separated by the outer glass wall of the capillary with a thickness of $h=20\pm2$~$\mu$m (CM Scientific). A dichroic mirror with a reflectivity $> 99\%$ at $\lambda=532$nm (for incident angles $0^\circ$ to $45^\circ$, custom made by Asphericon GmbH, Jena, Germany) is placed on the opposite side of the cell. The  mirror is transparent for wavelengths above $\lambda \sim650$~nm allowing both the trapping and the visual observation of the melamine microspheres. The clear layer sandwiched between the turbid layer and the mirror (cavity) thus has a thickness of $h+t\simeq35~\mu$m. The thin slab geometry together with the scattering in the turbid layer and multiple reflections in the cavity assure that in the center of the sample cell we generate a fairly isotropic and quasi-monochromatic random light field. The volume speckle inside the cavity displays intensity fluctuations on a typical length scale\cite{Emiliani2003} of $\lambda/2\sim~200$nm. In contrast with the anisotropic static light speckle pattern used to generate random potentials landscapes in the quest for Anderson localization of matter waves \cite{Billy2008}, we generate random temporal fluctuations of the light fields by rapidly scanning the green laser focal spot over the surface of the turbid layer. We set the scan distance to $y\pm 7~\mu$m at at an oscillation frequency of 500~Hz using a galvano mirror. The oscillation frequency is chosen in a way that light field fluctuates randomly much faster than $\tau_B=R^2/(6D_0)=776$~ms, the time scale for Brownian motion of the smallest melamine microspheres under study. Here $D_0$ denotes the Brownian diffusion coefficient. It has been recently reported that the collective motion of a large set of microspheres under the influence of both  Brownian and self-interacting optical forces becomes active and their dynamical quantities are no longer representative of thermodynamic equilibrium \cite{Douglass2012}. In our case, however, the rapidly fluctuating speckle is not coupled to the motion of the melamine spheres. It is important to note that roughly $2/3$ of the incident laser light will be reflected by the turbid layer. However, the reduced incident power is compensated by multiple reflections inside the cavity except for the small residual losses by the dichroic mirror. Therefore we expect the laser power in the cavity to be approximately the same as the incident laser power $P$. Reflections in the cavity can lead to an increase of the effective surface area $A$ as the reflected light can spread out laterally. To obtain an estimate of $A$ we image the residual green light, for an incident laser power of 1 W, transmitted by the dichroic mirror by increasing the exposure time and the gain setting of the camera corresponding to an increase in detection efficiency by more than four orders of magnitude. From an analysis of the slightly elliptical intensity distribution we derive an areal cross section of $A\sim \pi \times 41\times34\mu$m$^2\sim 0.004$mm$^2$ based on the $1/e$ decay length along the major and the minor axis. This means that for a laser power of up to 5 W we can indeed reach intensities $\sim$mW/$\mu$m$^2$  comparable to the case of optical tweezing. 

In one experimental run we observe the Brownian motion of two micron sized melamine microspheres that are held at a mean distance $r_0$ in the center of the light filled cavity, both laterally and axially, using a specifically adapted Nikon Eclipse TS100 bright field microscope composed of (i) a long working distance objective (20x/0.42 EO Plan Apo ELWD) for the sidewise white light illumination (ii) an oil immersion objective for both the trapping and the observation of the particle motion using a CCD camera and (iii) a notch filter to filter out residual stray light coming from the trapping laser as well as a dielectric mirror to couple the trapping laser into the optical path of the microscope. Note, the bright field illumination takes place across the first diffusing layer. For the laser trapping we use a near infrared laser beam (Toptica DL 100) with a wavelength of 785~nm and a power of $\sim$8~mW (measured at the exit of the laser) that is rapidly switched between two positions with a galvano mirror to produce the time shared dual-traps \cite{Sasaki1991, Fallman1997, Mio2000}.

Using a digital camera (Prosilica GC650, Allied Vision Technologies GmbH, Germany) we record images of 120~$\times$~120 pixels with a frame rate of 90~Hz and an exposure time of 0.3~ms at  $80 \kern 1 pt\text{X}$ magnification. The edge length per pixel is $d_{pix}\approx100~$nm which provides a sub-pixel localization accuracy of the particle center better than $10$nm \cite{Polin2008}. For a given mean separation distance $r_{0}$ we perform two experiments (see Figure \ref{fig:sample_cell}). In a first experiment we acquire a movie of 4000 images at a frame rate of 90~Hz under the influence of a random light field. In a subsequent reference experiment the random light field is turned off and the measurement is repeated under otherwise identical conditions. The recorded sequence of images is analysed using a standard particle tracking algorithm \cite{Crocker1996} to obtain distributions of the center to center separations of the trapped particles for both experiments. We follow the approach of Grier and coworkers \cite{Polin2008} to obtain an autocalibrated measurement of the colloidal interaction potential by analysing the differences between the distributions in the presence and absence of optically induced forces.

\section{Supplementary Material: Experiment}
\label{ExpSuppl2}
\subsection{Laser trapping experiment and data treatment}

The focused beam of a Topica DL 100 diode laser operating at a wavelength of 785~nm is time shared between two points using a galvano mirror (Galvoline G1432) driven by a square-wave oscillation at a frequency of 500~Hz. A telescope (Thorlabs, 3x Galilean optical beam expander, BE03M-B) is used to match the beam size with the back aperture of an oil immersion objective (Nikon 60x PlanApoVC, N.A.$=1.4$)\cite{Ashkin1992}. The location of the telescope is chosen in a way that the back focal plane of the oil immersion objective is imaged onto the galvano mirror, which allows for identical dual-traps\cite{Fallman1997}. Finally the time shared beam is focussed into the water layer of the sample cell to form the dual-trap. The particles are trapped in the middle of the water layer to minimize wall effects. The average distance between traps' centers can be changed by adjusting the amplitude of the galvano mirror oscillations. With the CCD camera we record images of 120~$\times$~120 pixels with a frame rate of 90~Hz and an exposure time of 0.3~ms. With a micro-scale the effective pixel size is measured to be $d_{pix}\approx0.1~\mu$m. The recorded images are analysed using an adapted MATLAB (The MathWorks, Inc., USA) code based on the particle tracking algorithm by Crocker and Grier\cite{Crocker1996} to finally obtain the particle positions on each picture. We quantify the transversal instrumental resolution of our apparatus by tracking two $2R=2~\mu$m particles with a center-to-center separation of $r_s$ that are permanently adsorbed to the lower glass surface of the water layer. A Gaussian fit to the measured  distribution reveals a standard deviation of $\sigma_{xy}=0.077$~pixel$~=7.7$~nm reflecting the transversal instrumental resolution. Moreover we have verified that out-of-plane fluctuations due to the finite trapping strength are negligible in our experiment \cite{Biancaniello2006}.

\begin{figure}[t]
\begin{center}
\includegraphics[width=10.0cm]{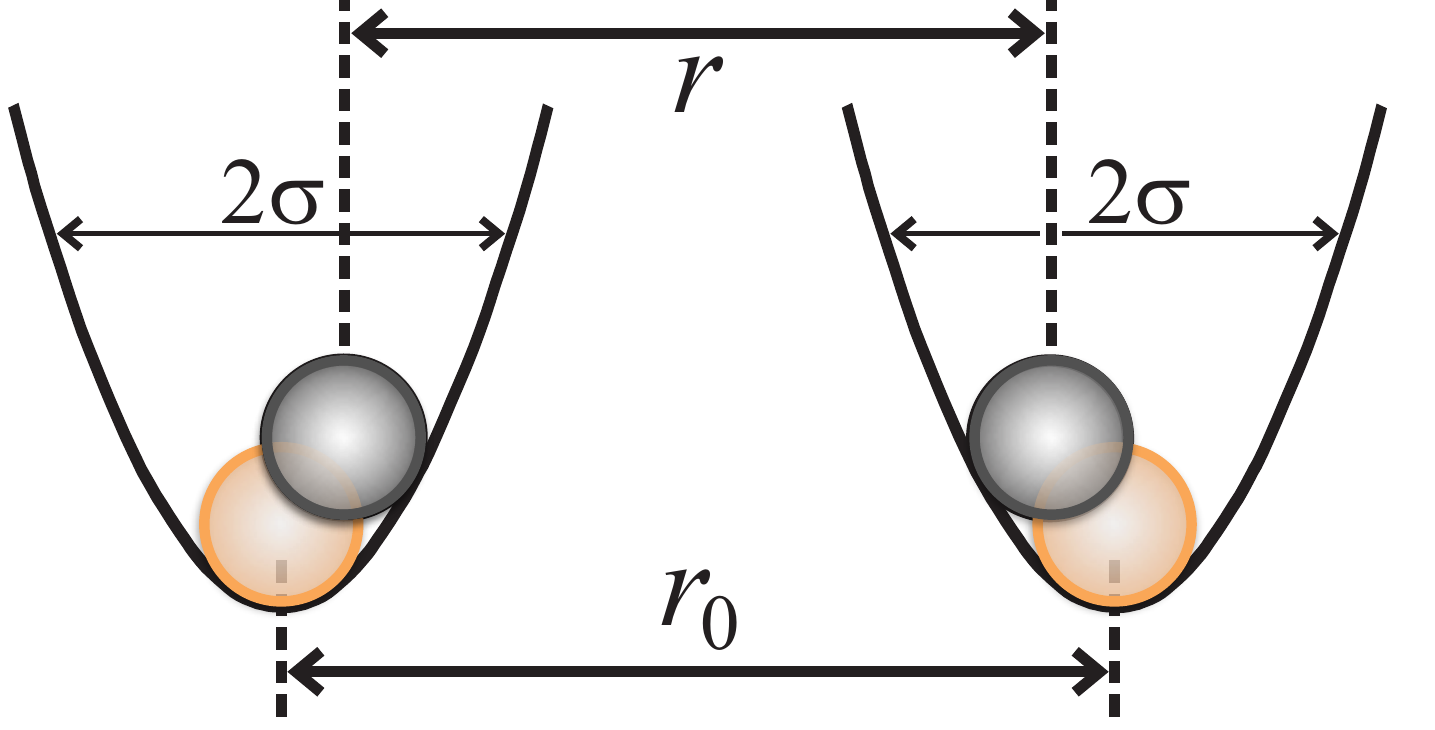}
\caption{{\small {\bf Thermal motion inside adjacent optical traps}.  Two particles with radius $R$ inside two identical optical traps are positioned at a distance $r_0$. For an isolated particle the distribution of positions is Gaussian with a standard deviation $\sigma$, which is set by the trap stiffness. Thus $2\sigma$ is a measure for the typical distances $r-r_0$ probed by thermal motion. Interactions between particles lead to a characteristic change in the distribution of particle positions. Attractive interactions increase the probability for the particles to approach. Precise measurements of $f_{pair}(r)$ are therefore a sensitive tool to determine the particle-particle interaction potential $U(D=r-2R)$.}}
\label{fig:procedure}
\end{center}
\end{figure}

The thermal motion of the two particles in the adjacent traps is illustrated in Figure \ref{fig:procedure}. For a given mean separation distance $r_{0}$ of the time shared optical traps we perform two experiments (see Figure~1 of main text).  In a first experiment we acquire a movie of 4000 images at a frame rate of 90~Hz under the influence of a random light field. In a subsequent reference experiment the random light field is turned off and the measurement is repeated under otherwise identical conditions. The recorded sequence of images is analysed using a standard particle tracking algorithm \cite{Crocker1996} to obtain distributions of the center to center separations of the trapped particles for both experiments. We follow the approach of Grier and coworkers \cite{Polin2008} to obtain an autocalibrated measurement of the colloidal interaction potential by analysing the differences between the distributions in the presence and absence of optically induced forces.

\begin{equation}
\label{eq:Url_RG}
 \frac{U\left( r \right)}{\mathrm{k}_BT} = \ln \left[ f_0  \left( r \right) \right] - \ln \left[ f_{rl}  \left( r \right) \right]
\end{equation}
where $f_{rl}$ and $f_0$ are the corresponding distributions of center-to-center separations. 
\newline From $n$ measurements of the  center-to-center separation $r$ we compute the pair distribution $f_{pair}(r)$ using the technique of nonparametric density estimation\cite{Silverman1986, Polin2008}:
\begin{equation}
\label{eq:density_estimation}
	f_{pair}\left(r\right) = \frac{1}{n h_{opt}}\sum_{i=1}^{n} J \left( \frac{r-r_i}{h_{opt}} \right)
\end{equation}
where $r_i$ reflects the separation distance determined from one image $i$ (one measurement); $h_{opt}$ is a smoothing parameter. The estimator's kernel $J[(r-r_i)/h_{opt}]$  can be any smooth function that satisfies the following conditions: (i) continuous and symmetric around zero (ii) integrable with its maximum $J_{max}$ at zero and (iii) normalized and non-negative \cite{Silverman1986}. For convenience we choose a Gaussian function of the form:
\begin{equation}
	J \left( \frac{r-r_i}{h_{opt}} \right) = \frac{1}{\sqrt{2\pi}} \exp \left[ \frac{\left( r-r_i \right)^2}{2 h_{opt}^2} \right]
\end{equation} 
The smoothing parameter $h_{opt}$ reflects the kernel's bandwidth. A proper choice of $h_{opt}$ is crucial. A too large width obscures features in the pair distribution $f_{pair}(r)$ whereas a too small width yield noisy results. A good trade-off is given by Silverman's rule \cite{Silverman1986}: $h_{opt}=\left(\frac{4}{3 n} \right)^{1/5}\sigma_r$, where $\sigma_r$ is the standard deviation of all separation distances $r_i$. The benefit of nonparametric density estimation over histograms is (i) the convergence speed; for $n$ data points the statistical error in histograms decreases as $n^{-1/2}$ whereas for nonparametric density estimation the error improves as $n^{-4/5}$ (see Refs.~\onlinecite{Thompson1990, Polin2008}). More importantly the nonparametric density estimation does not rely on the choice of discrete bins.

\subsection{Total interaction potential of the charge stabilized microspheres}
\begin{figure}[t]
\begin{center}
\includegraphics[width=10.0cm]{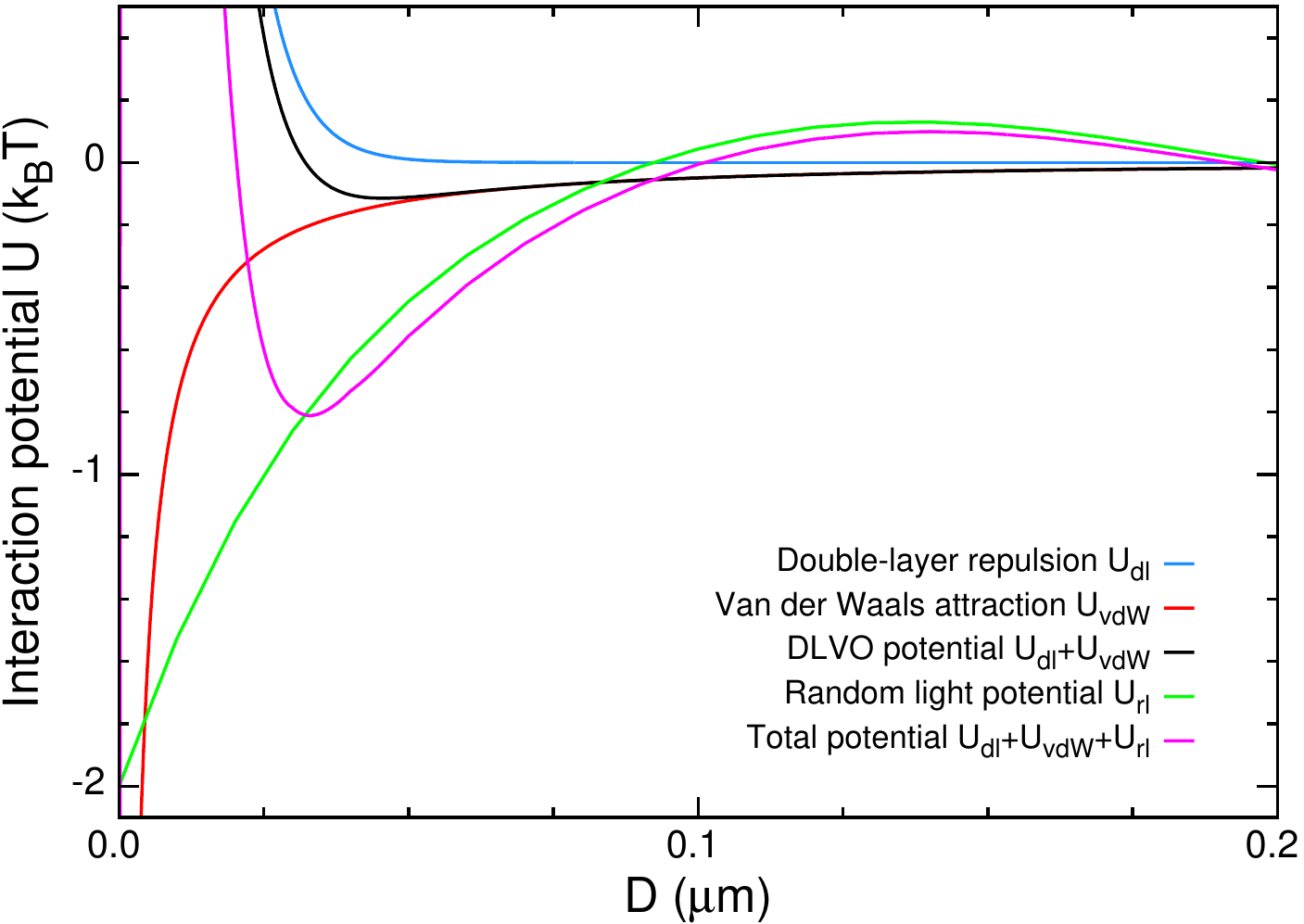}
\caption{{\small Calculated interaction potential for charge stabilized colloidal particles of size $2R=2\mu$m suspended in water in the presence and the absence of the random light (rl) induced interactions. The combination of van der Waals attractions (vdW - red line), Hamaker constant $A=0.1$~k$_B$T, and electrostatic double layer repulsions (dl-blue line), Debye length $\lambda_D=6$~nm, contact potential $U_{dl}(D=0)=44$~k$_B$T leads to the DLVO potential (black line). The attractive potential due to random light forces (green line), monochromatic random illumination with a contact potential of -2~k$_B$T, Eq.(6) is superimposed  leading to the total potential (magenta line). The repulsive part dominates at very short distances and in turn this allows us to probe the superimposed light induced attractions, without particles sticking together irreversibly.}}
\label{fig:DLVO}
\end{center}
\end{figure}

Particles suspended in water involve both van der Waals and double layer (dl) electrostatic repulsive interactions which in combination can be described by the well known DLVO theory \cite{Israelachvili1991}. The latter is dominantly repulsive for stable suspensions and thus prevents particle coagulation. Equally, in our measurements, this repulsive part dominates at very short distances and in turn this allows us to probe the superimposed light induced attractions, without particles sticking together irreversibly. For illustration we show in Figure \ref{fig:DLVO} a typical DLVO potential for micron sized particles ($2R=2\mu$m) and the superimposed attraction due to random light fields corresponding to the case shown in Figure 3 (a) in the main text. Exact values for the Hamaker constant and the contact potential $U_{dl}(D=0)$ are not known and we have chosen reasonable estimates consistent with the observed stability of the melamine particles. 

\subsection{Amorphous turbid layer}
The turbid layer at the entry of the light filled cavity is composed of a dense amorphous assembly of PMMA (Polymethylmethacrylate) particles, diameter $\sim 0.4\mu$m. We prepare the sample by filling a hollow rectangle borosilicate glass capillary (CM Scientific) with a height of $\simeq 20$~$\mu$m and a width of $w=200$~$\mu$m with a concentrated colloidal suspension with a particle volume fraction of approximately $\phi \approx 0.35$ and then let the suspension dry. We perform scanning electron microscopy on the dry particle layer by breaking the capillary after the experiment. The images (not shown) reveal a densely packed random structure in the bulk of the dried sample and a thin boundary layer with a crystalline structure close to the cell wall. We measure the line-of-sight transmission by collimating the 532nm laser beam and masking it with a 50~$\mu$m pinhole that we place as close as possible (ca. 1~mm) to the front surface of a glass capillary. On the opposite side we record the far field intensity profile by directly placing the sensor of the digital camera.  We estimate the line of sight transmission to $T_{los}=0.1$~\%. We estimate the total transmission by placing a high numerical aperture objective (Nikon 60x PlanApoVC, N.A.= 1.4) on the opposite side of the sample cell in order to maximise the acceptance angle for transmitted light. The collected light is then projected on a screen, imaged and analyzed with the digital camera.  From this we obtain an estimate for the diffuse total transmission of $T_{diff}\sim1/3$.

\section{Supplementary Material: Theory}
\label{TheorSuppl} 

%
%
%
\subsection{Field-Field correlations and cross-spectral density in a stationary random field}
We consider a fluctuating electric field, $\E(\r,t)$  in a transparent and non-dispersive homogeneous medium with real refractive index $n_h=\sqrt{\epsilon_h}$.
For a stationary field\cite{Joulain2005}, the spatiotemporal fluctuations,
\be
\langle E_i(\r,t) E_j(\r',t') \rangle = \text{Re}\left\{ \int_{0}^{\infty} \frac{d\omega}{2 \pi} 
W_{ij}(\r,\r', \ \omega) e^{-i\omega(t-t')} \right\},
\ee
are characterized by the the cross-spectral density tensor
 $ W_{ij}(\r,\r', \ \omega) $  given by \cite{Setala2003}
 \be
 W_{ij}(\r,\r', \ \omega) &\equiv& \langle E^*_i(\r,\omega) E_j(\r',\omega) \rangle = \frac{4 \pi} {\epsilon_0 \epsilon_h}  u_E(\r, \omega)  \  \left\{\frac{2 \pi}{k}\text{Im}\{G_{ij}(\r,\r', \ \omega)\}\right\}  \ee
 where $G_{ij}(\r,\r', \ \omega)$ are the matrix elements of the  Green tensor,
 \begin{eqnarray}
 \G(\r,\r')  = \frac{k}{4\pi} \left [ {\bf{I}}_3+\frac{1}{k^2}\bm{\nabla}\bm{\nabla} \right ] \, \frac{\exp(i k |\r-\r'| R)}{k |\r-\r'|} \label{eq:G},
\end{eqnarray}
(${\bf I}_3$ is the identity tensor) and 
 $u_E(\r, \omega)$ is related to the time-averaged electric energy per unit volume, 
 \be \langle U(\r,t) \rangle = \frac{\epsilon_h \epsilon_0}{2} \langle | \E(\r,t)|^2 \rangle = 
  \int_0^\infty u_E(\omega) d\omega. \ee 

\subsection{Multiple scattering between two compact bodies}

We consider a particle, $A$ 
centered at the origin of coordinates and a particle $B$  displaced a distance $r$ along the positive $z$-axis. From a physical point of view, instead of a continuous approach, 
each particle can be seen as made of discretized, $N_{A}$ and  $N_{B}$, identical cubic elements   of volume $v$. This is also known as a discrete dipole approach (DDA) \cite{Chaumet2000DDA}.
In the presence of an external polarizing field, $\E_{\text{inc}}(\r,\omega)$, each volume element acts as an induced dipole proportional to the polarizing field, i.e. \be
 \p(\r_n, \omega) = \epsilon_0 \epsilon_h  \ \alpha(\omega) \E_{\text{inc}}(\r_n, \omega)  \label{pdef} \ee
 where $\alpha(\omega)$ is the polarizability  which, for cubic or spherical elements of volume $v$, is given by\cite{Albaladejo2010}
\be
\alpha(\omega) \equiv  \frac{v \alfa_0}{1-i\frac{v k^3}{6\pi}\alfa_0} \quad, \quad \alfa_0(\omega) \equiv 3 \frac{\epsilon(\omega)-\epsilon_h}{\epsilon(\omega) + 2\epsilon_h}. \label{alfa}
\ee
The polarizing field on a given element in particle $B$,  $ \E_{\text{inc}}(\r_n^B, \omega)$, is  given by the solution of the multiple scattering problem:
\be
\E_{inc}(\r_n^B) &=&  \E_0(\r_n^B) +  \alpha_B k^2  \sum_{m \ne n}^{N_B} \G(\r_n^B,\r_m^B) \E_{inc}(\r_m^B) 
+ \alpha_A k^2  \sum_{m}^{N_A} \G(\r_n^B,\r_m^A) \E_{inc}(\r_m^A)    \label{DDA1} \ee
(and an equivalent equation for particle $A$). For simplicity in the notation, we do not include here the explicit $\omega$-dependence.
These are a set of $3N_A +3N_B$ equations that can be written in compact matrix form as
\be
 \E_{\text{inc}}(B) &=& \E_0(B) + \alpha_B k^2 \G_{B,B} \E_{\text{inc}}(B) + \alpha_A k^2 \G_{B,A} \E_{\text{inc}}(A) \\
  \E_{\text{inc}}(A) &=& \E_0(A) + \alpha_A k^2 \G_{A,A} \E_{\text{inc}}(A) + \alpha_B k^2 \G_{A,B} \E_{\text{inc}}(B) \ee
Introducing the $T$-matrix, defined as
\be
{\bf{T}}^{-1}(\r_n^B,\r_m^B) = \frac{1}{\alpha_B k^2} {\bf{I}}_{3}- \G(\r_n^B,\r_m^B) (1-\delta_{nm}),
\quad \text{or} \quad {\bf{T}}_B ^{-1} \equiv \frac{1}{\alpha_B k^2} {\bf{I}}_{3N_B} -   \G_{B,B}
\ee
where ${\bf{I}}_{3N_B}$ is the $3N_B \times 3N_B$  identity matrix (and an equivalent expression for ${\bf{T}}_A$),
the formal solution of the scattering problem can be written as 
\be
 \E_{\text{inc}}(B)&=& \frac{1}{\alpha_B k^2} \left( {\bf{T}}_B ^{-1}- \G_{B,A}{\bf{T}}_A\G_{A,B}\right)^{-1} \   \Big( \E_0(B) +  \G_{B,A} {\bf{T}}_A \E_0(A)\Big) \\
 \E_{\text{inc}}(A)&=&\frac{1}{\alpha_A k^2}  \left( {\bf{T}}_A ^{-1}- \G_{A,B}{\bf{T}}_B\G_{B,A}\right)^{-1} \  \Big( \E_0(A) +  \G_{A,B} {\bf{T}}_B \E_0(B)\Big) 
 \ee
 Our approach can be seen  as the  DDA-like version of the well known $T$-matrix approach of multiple scattering of electromagnetic waves by two different objects  usually described in terms of a basis of multipolar vector wave functions (see for example Ref. \cite{mishchenko2002scattering,waterman1971}).

 \subsection{Optical interactions between two compact bodies induced by random light fields}
 
 In the presence of a random (stationary) field, $\E_0(\r,t)$, both the dipoles
and the polarizing fields are, in general,  fluctuating quantities and the time averaged force along the $z$-axis may be
written as the sum of two different terms [see for example, Ref. \cite{henkel2002}] 
 \be
 F_z = \biggl\langle \p^{\text{ind}}(\r_n,t) \left. \frac{\partial}{\partial z} \E_{\text{inc}}^{\text{fluc}}(\r,t)\right|_{\r=\r_n} \biggr\rangle
+ \biggl\langle \p^{\text{fluc}}(t) \left. \frac{\partial}{\partial z} \E_{\text{inc}}^{\text{ind}}(\r,t)\right|_{\r=\r_n}
\biggr\rangle \label{fop}
 \ee
where the first term describes the force induced by the
fluctuating (external) field, $\E_0^{\text{fluc}}$ with the
corresponding induced dipole $\p^{\text{ind}}$ as discussed in the main text. The second
involves the  (spontaneous and thermal) fluctuations of the dipole
$\p^{\text{fluc}}$.  

We focus on lossless particles and discard the second contribution  (in absence of absorption, there are no spontaneous and thermal fluctuations of the dipoles). 
From Eqs. \eqref{pdef} and \eqref{fop}, the total time-averaged force on particle $B$ is then given by
\be
F_z^B &=& \epsilon_0 \epsilon_h \text{Re}\left\{ \int_0^\infty \frac{d\omega}{2\pi} 
  \alpha_B(\omega)  \sum_n^{N_B}  \biggl\langle \E_{\text{inc}}(\r_n^B,\omega) . \left. \frac{\partial}{\partial z} \E^*_{\text{inc}}(\r,\omega)  \right|_{\r=\r_n^B} \biggr\rangle \right\}  \ee
where $\E_{\text{inc}}(\r_n,\omega)$ is given in   \eqref{DDA1} and the gradient of the incoming field is the sum of three different terms
\be
\left. \frac{\partial}{\partial z} \E^*_{\text{inc}}(\r,\omega)  \right|_{\r=\r_n^B} &=&  \left. \frac{\partial}{\partial z} \E_0(\r) \right|_{\r=\r_n^B}+ \alpha_A k^2  \sum_{m}^{N_A}  \left.\frac{\partial}{\partial z}\G(\r,\r_m^A) \E_{inc}(\r_m^A)  \right|_{\r=\r_n^B}   \nonumber \\ && 
+  \alpha_B k^2  \sum_{m \ne n}^{N_B}  \left.\frac{\partial}{\partial z}\G(\r,\r_m^B) \E_{inc}(\r_m^B) \right|_{\r=\r_n^B}\label{DeDDA1} \ee
These three terms give three different contributions to the total force on particle $B$.
 The first term, $F_z^{B1}$, can be seen as 
coming from the homogeneous radiation field on particle $B$ (which after arriving at $B$, suffers multiple scattering events with particle $A$). The second, $F_z^{B2}$, comes from the radiation  first scattered by particle $A$. The last term, arising from the multiple scattering interactions inside the particle, does not contribute to the total force on $B$ since these interactions cancel out  when summing over all the dipoles in $B$ after averaging over the random field. 
Taking into account that 
\be
\biggl\langle E_{0j}(\r_n^A,\omega)  \left. \frac{\partial}{\partial z} \E^*_{0i}(\r,\omega)  \right|_{\r=\r_m^B} \biggr\rangle &=&   \frac{u_E( \omega)}{\epsilon_0 \epsilon_h}  \frac{8 \pi^2}{k}\text{Im}\left( \frac{\partial}{\partial z}\{G_{ij}(\r,\r_n^A, \ \omega)\} \right)_{\r=\r_m^B}  \\
\left( \frac{\partial}{\partial z}\{\G(\r,\r_n^A, \ \omega)\} \right)_{\r=\r_m^B} &=& 
\frac{\partial}{\partial r}\{\G(\r_m^B-\r_n^A, \ \omega)\}
\ee
we find $F_z^B = \int_0^\infty d \omega[F_z^{B1}(\omega) +  F_z^{B2}(\omega)]$ with
\be
F_z^{B1}(\omega) &=&  \frac{4 \pi u_E( \omega)}{k^3} 
 \text{Tr} \left[
\text{Im}\left\{ \frac{\partial}{\partial r} \G_{B,A}\right\}  \text{Re} \left\{{\bf{T}}_A \G_{A,B}
\left( {\bf{T}}_B ^{-1}- \G_{B,A}{\bf{T}}_A\G_{A,B}\right)^{-1}\right\} \right]
\label{FB1} \ee
where ``Tr'' stands for the trace of the $3N_B\times 3N_B$ matrix.
After some algebra and taking into account that 
in absence of absorption (i.e. $\epsilon(\omega)$ and $\alfa_0$ are real) 
\be
\text{Im}{\bf{T}}_B ^{-1} \equiv \frac{k}{6\pi} {\bf{I}}_{3N_B} -   \text{Im}\G_{B,B} \quad, \quad  
\text{Im}{\bf{T}} ^{-1}(\r_n^B,\r_m^B) = - \text{Im}\G(\r_n^B,\r_m^B) ,
\ee
the second contribution can be shown to be given by
\be
F_z^{B2}(\omega) &=&  \frac{4 \pi u_E(\omega)}{k^3} 
 \text{Tr} \left[
\text{Re}\left\{ \frac{\partial}{\partial r} \G_{B,A}\right\}  \text{Im} \left\{{\bf{T}}_A \G_{A,B}
\left( {\bf{T}}_B ^{-1}- \G_{B,A}{\bf{T}}_A\G_{A,B}\right)^{-1}\right\} \right]
\label{FB2}\ee
Adding \eqref{FB1} and \eqref{FB2} we finally obtain that, in absence of absorption, 
the total force is conservative
$
F_z^B = - \partial U(r)/ \partial r$ with an interaction potential given by 
\be
U(r) =  \int_0^\infty d \omega \ 
\frac{2 \pi}{k^3} u_E(\omega) \ \text{Im}  \text{Tr} \left[\ln\left(  {\bf{I}} - \G_{B,A}{\bf{T}}_A\G_{A,B}{\bf{T}}_B \right) \right] \label{Uexact2}
\ee
The dependence of the interaction on distance is completely contained in $\G_{B,A}$ whereas all the shape and material dependence is contained in the $T$ matrices. 
 
In the case of  equilibrium thermal blackbody radiation  the electric energy density, $U_E(\omega) $, is given by \cite{Landau_Statistical,Joulain2005}
\be
 u_E(\omega)d\omega =  \frac{\hbar \omega}{2} \coth\left( \frac{\hbar \omega}{2K_BT}\right)  \frac{n_h^3 \omega^2}{2 \pi^2 c^3} d\omega \ee
 which, at  zero temperature gives $u_E(\omega) = \hbar k^3/(4\pi^2)$. For absorbing (emitting) particles in equilibrium, we can include in Eq. \eqref{fop}  the contribution of the fluctuating dipoles  and the corresponding radiated fields \cite{henkel2002} (linked through the fluctuation-dissipation theorem). Interestingly, in equilibrium, this additional contribution conspires with the force due to the field fluctuations  to give a total interaction potential which is exactly given by  Eq. \eqref{Uexact2}, now including light absorption and emission (i.e. [$\text{Im}\{\epsilon(\omega)\} \ge 0$) and  we recover the exact Casimir interaction between arbitrary compact objects \cite{kenneth2006,Emig2007}.

{
\subsection{Attractive and repulsive interactions between dipolar electric and magnetic  particles} 

Submicron dielectric spheres made of moderate permittivity materials present dipolar magnetic and electric responses \cite{Garcia_Etxarri2011}, characterized by their respective first-order ``Mie'' coefficients, in the near infrared, in such a way that either of them can be selected by choosing the illumination wavelength.  The scattering properties of Silicon and other semiconductor nanoparticles  \cite{Garcia_Etxarri2011},  can be well described by their electric and magnetic polarizabilities, being negligible  the contribution of higher order modes (contribution of higher order modes can be relevant when the interparticle distance $D$ becomes of the order of the particle radius  -a detailed analysis of these interactions will be described elsewhere-).  
 
When the optical response of the particles can be described by their electric and magnetic polarizabilities,  $\alpha_n^e(\omega)$ and $\alpha_n^m(\omega)$ respectively ($n=A,B$). The presence of an external polarizing field  induce both electric, $\p$ and magnetic, $\m$, dipoles, i.e. 
\be
 \p(\r_n, \omega) &=& \epsilon_0 \epsilon_h  \ \alpha_n^e(\omega) \E_{\text{inc}}(\r_n, \omega)  \\
 \m(\r_n, \omega) &=&    \alpha_n^m(\omega) \H_{\text{inc}}(\r_n, \omega) =  - 
 \frac{i \alpha_n^m(\omega) }{kZ} \left.  \bm{\nabla} \times \E_{\text{inc}}(\r, \omega) \right|_{\r = \r_n}
 \ee
  where  $ Z \equiv \sqrt{\mu_0/(\epsilon_0 \epsilon_h)}$ is the  impedance of the homogeneous medium. The polarizabilities are simply related to the first electric, $a_1$, and magnetic, $b_1$ Mie coefficients:
 \be
 \alpha_n^e(\omega) = i \frac{6\pi}{k^3} a_1 \quad,\quad \alpha_n^m(\omega) = i \frac{6\pi}{k^3} b_1.
 \ee
 
 When we just consider dipolar particles we can write 
\be
\quad {\bf{T}}_B = k^2 \begin{pmatrix}
\alpha_B^e {\bf I}\hfill & 0 \\
0 & \alpha_B^m {\bf I}\hfill 
\end{pmatrix}
= \begin{pmatrix}
T_B^e {\bf I} \hfill & 0 \\
0 & T_B^m {\bf I} \hfill 
\end{pmatrix}
\quad \text{and} \quad
 \G_{BA} \equiv \begin{pmatrix}
 \G_E(B,A) \hfill &   \G_M(B,A) \\
   \G_M(B,A) &   \G_E(B,A) \hfill 
\end{pmatrix} \label{GBA}
\ee
with
 \be
\G_E(B,A) = \begin{pmatrix}
G_{E,x} \hfill & 0 \hfill & 0 \\
0 \hfill & G_{E,x} \hfill & 0 \\
0 \hfill & 0 \hfill & G_{E,z} 
\end{pmatrix} \quad, \quad  \G_M(B,A) = \begin{pmatrix}
0 \hfill & -G_{M} \hfill & 0 \\
G_{M} \hfill & 0 \hfill & 0 \\
0 \hfill & 0 \hfill & 0 
\end{pmatrix}
\ee
being  \be
 G_{E, x}(r)&=&  G_{E, y}(r) = \left( 1 + \frac{i}{kr}-\frac{1}{k^2 r^2} \right)g(r) \\
 G_{E, z}(r)&=&   \left( - \frac{2i}{kr}+\frac{2}{k^2 r^2} \right)g(r) \\
 G_{M}(r) &=& \left( i - \frac{1}{kr} \right) g(r).
 \ee
 and $g(r)= e^{ikr}/(4\pi r)$ the scalar Green function.
 
In absence of absorption the trace formula for the interaction potential   [Eq. \eqref{Uexact2}] can be calculated in closed form as:
\be
  \text{Tr} \left[\ln\left(  {\bf{I}} - \G_{B,A}{\bf{T}}_A\G_{A,B}{\bf{T}}_B \right) \right]  &=& \nonumber  \\
= \ln\Big(1-T_B^eT_A^e G_{Ez}^2(r)\Big) &+&\ln\Big(1-T_B^mT_A^m G_{Ez}^2(r)\Big)  \nonumber \\
+ 2 \ln\Bigg[ \Big(1-T_B^e\left(T_A^e G_{Ex}^2(r) +T_A^m G_M^2(r)\right) \Big)&&
\Big(1-T_B^m\left(T_A^m G_{Ex}^2(r)+T_A^e G_M^2(r)\right) \Big) \nonumber \\ &&-T_B^eT_B^m\left( T_A^m - T_A^e\right)^2G_{Ex}^2(r)G_M^2(r) \Bigg]
\ee
As shown in Figure 2 in the main text, for two identical particles near the first Mie dipolar magnetic resonance  the interaction force can be repulsive in analogy with the repulsive interactions between resonant atoms \cite{Walker1990}.
 }

\subsection{Weak scattering approximation}

In the weak scattering limit,  we can expand \eqref{Uexact2} 
\be
U(r) &\approx& - \int_0^\infty d \omega \ 
 u_E(\omega) \ 2 \pi k \ \alfa_0^2 v^2 \text{Im} \ \text{Tr} \left[ \G_{B,A}\G_{A,B}  \right] + {\cal{O}}(\alfa_0^3) \\
&=& - \int_0^\infty d \omega  \ 
 u_E(\omega) \ 2 \pi k \ \alfa_0^2 v^2 \sum_{n}^{N_B} \sum_{m}^{N_A} \sum_{i,j} \text{Im} \left[ G_{ij}^2(\r_n^B-\r_m^A)  \right] \\
&=& - \int_0^\infty d \omega  \ 
 u_E(\omega) \ 2 \pi k \ \alfa_0^2 \int_B d\r_B^3 \int_A d\r_A^3  \sum_{i,j} \text{Im} \left[ G_{ij}^2(|\r_B-\r_A|)\right].
\ee
It is easy to see that this is the result obtained at lowest order in the so-called Born (or Rayleigh-Gans-Debye) approximation.
For two identical spheres of radius $R$, their centers being a distance $r$ apart,   in a quasi monochromatic random field,  the interaction energy in the weak scattering
limit can be  seen as a Hamaker's integral \cite{Hamaker1937}  
{
 \be 
U\left( D \right) &=&   -K \times {\cal U}(D,R,k) \nonumber \\
{\cal U}(D,R,k) &=& \frac{\pi ^2}{r}\int\limits_{r - R}^{r + R} {dy \left[ \left( {{R^2} - {{\left( {r - y} \right)}^2}} \right)\int\limits_{y - R}^{y + R} {dx} \Big\{ {\left( {{R^2} - {{\left( {y - x} \right)}^2}} \right)x\,f\left( x \right)} \Big\} \right]}  \label{JuanjoRGD} \\
f\left( x \right) &=& 
 \left( \frac{4\pi}{k}\right)^2 \text{Im}\left\{ \sum_{i,j}  G_{ji}^2(x)\right\} \nonumber \\
 &=& \text{Im}\left\{  e^{2ikx} \left( \frac{2}{\left( {kx} \right)^2} + \frac{4i}{\left( {kx} \right)^3} - \frac{{10}}{\left( {kx} \right)^4} - \frac{{12i}}{\left( {kx} \right)^5} + \frac{6}{\left( {kx} \right)^6} \right) \right\} \label{Gij2}
\ee
 with $D=r-2R$ and $K=\{  d \omega \ 2
 u_E(\omega)\} \pi {k^3}\left(\alfa_0/(4\pi)\right)^2$.
}
%

{
\subsection{Random light forces between dipolar  electric particles:  gravitational like interactions}
 
 If we consider the long wavelength limit,  where the magnetic polarizability is negligible,
 Eq. \eqref{Uexact2} takes the simple form 
  \begin{align}
U(r) &=   
\frac{2 \pi}{k^3} U_E(\omega) \  \text{Im} \left\{\sum_{i=x,y,z}
 \ln\left( \left[1-  \left(\alpha^e k^2 G_{ii}(r)\right)^2\right]  \right) \right\} \label{UBoyer}
 \end{align}
 which can be shown to  be equivalent to Eq. (11) in Ref. \cite{Sukhov2013} in absence of absorption.

 A remarkable prediction concerning optically induced interactions between atoms, molecules or small dipolar particles  \cite{Thirunamachandran1980} is that,  after averaging over all orientations of the inter-atomic axis with respect to the incident beam,
the interaction is an isotropic long-range, ``gravitational-like", $1/r$  potential  in the near field. 
 It was suggested \cite{Odell2000} that this averaging  could be experimentally achieved by  an isotropic external illumination by means of multiple incoherent beams which, for atomic systems, could give rise to stable Bose-Einstein condensates with unique static properties \cite{Odell2000}. An alternative is to average over all orientations and polarizations of the incoming, uncorrelated, plane waves \cite{Sukhov2013}:
 In the weak scattering limit, expanding Eq. \eqref{UBoyer} leads to
  \begin{align}
U(r) &\approx   
- \frac{2 \pi}{k^3} U_E(\omega) \  \text{Im} \left\{\sum_{i=x,y,z}
 \left(\alpha^e k^2 G_{ii}(r)\right)^2 \right\} 
 \end{align}
 which in the short distance limit gives the
 above mentioned  long-range $1/r$ dependence of the optical interaction potential  
 in agreement with previous results \cite{Thirunamachandran1980,Sukhov2013}.
 It is worth noticing that similar ideas were considered in the earlier proposal by Spitzer \cite{Spitzer1941} of the so-called mock gravity, gravity-like interactions between matter in the universe due to background isotropic radiation pressure.}

\end{document}